
\input phyzzx
\hoffset=0.375in
\overfullrule=0pt

\def\dol{{D_{\rm OL}}}
\def\dls{{D_{\rm LS}}}
\def\dos{{D_{\rm OS}}}

\def\pc{{\rm pc}}

\def\kms{{\rm km}\,{\rm s}^{-1}}

\def\ms{{\rm Machos}\ }

\def\msk{{\rm Machos}}
\twelvepoint
\font\bigfont=cmr17
\centerline{}
\bigskip
\bigskip
\singlespace
\centerline{\bigfont Microlensing by Stars in the Disk of M31}
\bigskip
\centerline{\bf Andrew Gould}
\smallskip
\centerline{Dept of Astronomy, Ohio State University, Columbus, OH 43210}
\smallskip
\centerline{E-mail Gould@payne.mps.ohio-state.edu}
\bigskip
\vskip 1.7in
\centerline{\bf Abstract}

\singlespace

	The optical depth to microlensing toward M31 due to known stars
in the disk of M31 itself is $\tau\sim 2\times 10^{-7}e^{-r/d}$ where
$d$ is the disk scale length and $r$ is the distance along the major axis.
Thus, there can be significant lensing toward the M31 disk even if
M31 contains no dark compact objects.  The optical depth has a strong
dependence on azimuthal angle: at fixed radius
$\tau\propto[1+ (h/d)\tan i\cos\phi]^{-2}$ where $h$ is the scale height
of the disk, $i=75^\circ$ is the inclination of M31,
and $\phi$ is the azimuthal angle relative to the near minor axis.
By measuring the optical depth as a function of radial and azimuthal
position, it is possible to estimate $h$ and $d$ for the mass of the
M31 disk, and so determine whether the disk light traces disk mass.
Ground-based observations in $0.\hskip-2pt''5$ seeing of 0.8 $\rm deg^2$
once per week could yield $\sim 3$ events per year.
With an ambitious space-based project, it would be possible to
to observe $\sim 80$ events per year.
If lensing events were dominated by the M31 spheroid rather than the disk,
the event rate would be higher and the spheroid's parameters could be
measured.  The pattern of optical depths differs substantially for a disk
and a spheroid.

\noindent Subject Headings:  dark matter -- gravitational lensing --
galaxies: individual: M31
\vskip 0.3in
\centerline{submitted to {\it The Astrophysical Journal}: March 10, 1994}
\bigskip
\centerline{Preprint: OSU-TA-5/94}

\endpage

\chapter{Introduction}
\normalspace

	Three groups have recently detected a total of at least 13 candidate
microlensing events presumably caused by compact objects in the Galaxy which
magnify the light of distant stars (Alcock et al.\ 1993, MACHO;
Aubourg et al.\ 1993, EROS; Udalski et al.\ 1993, OGLE).
The observations by MACHO and EROS of several million stars toward the
Large Magellanic Cloud (LMC) were originally suggested by
Paczy\'nski (1986) as a way to detect massive compact halo objects (\msk)
that may make up the dark matter in galactic halos.  The observations by OGLE
and MACHO
toward the Galactic bulge were originally suggested by Paczy\'nski (1991) and
Griest et al.\ (1991) as a way to detect disk dark matter, to measure the
low-mass end of the disk luminosity function,
and better to constrain the distribution
of \msk.  However, even if some or all of the candidates prove to be real
lensing events, this will not necessarily imply that the lenses make up (or
even lie in) a dark halo.  Gould (1994) suggested that the low event rate
toward the LMC may indicate that the lenses lie in a dark thick disk.
Gould, Miralda-Escud\'e, \& Bahcall (1994) suggested that the lenses may
lie in a thin disk and devised several methods to distinguish among the
thin-disk, thick-disk, and halo models.  Gould et al.\ argued that the high
event rate toward the bulge compared to that toward the LMC tends to favor
a disk origin of the events. Giudice, Mollerach, \& Roulet (1994) suggested
that the MACHO and EROS events may be due to a dark spheroid, that is a
spherical structure that cuts off much more rapidly than an
$r^{-2}$ halo.

	Crotts (1992) pointed out that the optical depth toward the disk of M31
by \ms in an M31 halo would be far higher than the optical depth toward
the LMC by \ms in a Galactic halo.  He also noted that observations toward
M31 would have several other advantages, most notably that the near side
of the M31 disk would have a very much lower lensing rate and therefore
would provide an excellent control field.  The principal drawback of
observations toward M31 is that the fields are too crowded to resolve the
stars.  However, Tomany \& Crotts (1994) have pioneered a ground-based
method for detecting lensing events even for unresolved stars in $\gsim 1''$
seeing.  Moreover, future space-based observations with $0.\hskip-2pt ''1$
seeing would be able to resolve stars in M31 approximately as well as current
ground-based observations do for stars in the LMC.  Lensing searches toward
M31 are therefore technically feasible.

	Here I point out that even if there are no compact dark objects
in M31, the optical depth toward the M31 disk (due to known stars within the
M31 disk itself) is quite high, $\tau\sim
 10^{-7}$.   This evaluation holds at approximately one
disk scale length along the major axis of the disk.  The optical depth falls
directly as the disk density (i.e.\ exponentially) as one proceeds radially
outward.  For fixed radius, the optical depth varies strongly as a function
of azimuthal position.  By measuring these variations, it should be
possible to measure the scale height and scale length as well as the
normalization of the mass of the M31 disk.  These measurements would provide
a direct check of the frequently used (but never tested) hypothesis that
disk light traces disk mass.

	Of course, it is possible that the dark, $r^{-2}$ halo of
M31 is made of \msk.  In this
case, the microlensing due to low-mass stars in the M31 disk would be
dwarfed by that due to \msk, and it is unlikely that one could recognize
the characteristic signature of disk lenses.  Similarly, it is possible that
M31 has a dark spheroid (a roughly spherical structure with a steeper, e.g.,
$r^{-3.5}$ profile) which makes up some of the M31
dark matter, particularly
in its inner portions.  In this case also, it would be difficult to recognize
the signature of a disk.  However, in both these cases, it would be quite
easy to recognize that the signal was due to a halo or to a spheroid and
to map the characteristic features of either.  Thus, in any event, lensing
searches toward the M31 disk will yield valuable information about the
mass distribution of M31.

\chapter{Lensing By A Distant Disk}

	The Einstein radius, $r_e$ about a lens of mass $M$ has a
physical size given by
$$r_e^2 = {4 G M \dol\dls\over \dos c^2},\eqn\redef$$
where $\dol$, $\dos$, and $\dls$ are the distances between the observer,
source, and lens.  If both the lens and the source are in M31, then
$\dol\sim\dos$, so that
$$r_e^2 = {4 G M y\over c^2},\eqn\redeftw$$
where $y\equiv \dls$ is now the distance between the lens and source.

\section{Lensing of a Source in the Disk Plane}

	I assume that the mass of the M31 disk is distributed in a double
exponential profile,
$$\rho(r,z) = {\Sigma_0\over 2 h}\exp\biggl(-{|z|\over h}\biggr)
\exp\biggl(-{r\over d}\biggr),\eqn\rhoofrz$$
where $h$ and $r$ are respectively the scale height and scale length of the
disk and $\Sigma_0$ is the normalization.  The density along the line
of sight for a position $y\equiv\dls$ relative to a source at radius $r$
in the plane is then
$$\rho(y) = {\Sigma_0\over 2 h}\exp\biggl(-{y\cos i\over h}-{y\sin i \cos\phi
\over d}\biggr)\exp\biggl(-{r\over d}\biggr)\eqn\rhoofy$$
where $i=75^\circ$ is the angle of inclination of M31 and $\phi$ is the
azimuthal angle relative to the near minor axis.
The optical depth to a source in the plane of the M31 disk is therefore
$$\tau_0 ={ 2\pi G\Sigma_0 h\over c^2 }
\sec^2 i\biggl(1 + {h\over d}\tan i \cos \phi\biggr)^{-2}\
e^{-r/d}
\eqn\taueq$$

\section{Lensing of Exponentially Distributed Sources}

	Equation \taueq\ is appropriate if the majority of observed sources
are very young (e.g.\ blue supergiants)
and hence have a much lower scale height
than the lenses.  (The same result applies in the hypothetical case that
the lenses were confined to a plane and the sources were distributed
exponentially.)\ \  At the opposite extreme, one may assume that the sources
are distributed like the lenses, with exponential scale height $h$.  In this
case I find that the optical depth, $\tau_{\rm }$, is higher
$ \tau = {(3/ 2)}\tau_0$, so that
$$\tau(r,\phi) = { 3\pi G\Sigma_0 h\over c^2 }
\sec^2 i\biggl(1 + {h\over d}\tan i \cos \phi\biggr)^{-2}e^{-r/d}
.\eqn\tauexp$$

	For observations carried out at optical wavelengths, one
may expect that sources behind the M31 plane will be essentially completely
blocked by dust.  I then find that if the sources and lenses have the
same exponential scale height, the optical depth is lower,
$\tau = (1/2)\tau_0$,
$$\tau(r,\phi) = {\pi G\Sigma_0 h\over c^2 }
\sec^2 i\biggl(1 + {h\over d}\tan i \cos \phi\biggr)^{-2}e^{-r/d}
.\eqn\tauexptwo$$

	For realistic optical observations, I expect that the actual
optical depth will lie between $(1/2)\tau_0$ [eq.\ \tauexptwo] and
$\tau_0$ [eq.\ \taueq], and more likely closer to the former.  For
definiteness, and to be conservative, I henceforth compute optical
depths according to equation \tauexptwo.

\chapter{Mapping the M31 Disk Mass Distribution}

	From equation \tauexptwo\ the lensing rate along the major axis is
$$\tau = {\pi G \Sigma_0 h\over c^2}\sec^2 i\,e^{-r/d},
\qquad {\rm (major}\ {\rm axis)}. \eqn\taueqt$$

	I adopt a disk scale length, $d=6.4\,$kpc and total blue
luminosity $L=2.4\times 10^{10}\,L_\odot$ from van der Kruit (1989).
I somewhat arbitrarily adopt a mass-to-light ratio $M/L_B=3$ for the M31
disk, and a disk scale height $h=400\,$pc.  These values lead to an
estimate of $\Sigma_0= 280\,M_\odot\,\pc^{-2}$, and hence
$$\tau\sim 2.5\times 10^{-7}e^{-r/d},\eqn\mtoopt$$

	If the optical depth were measured at various positions
along the major axis, then from equation \taueqt\ one could determine
$d$ and $(\Sigma_0 h)$.  The radial dependence should be quite strong.
Then by measuring the azimuthal dependence
at fixed radius, one could determine $(h/d)$.  See equation \tauexptwo.
Note that for the adopted parameters, $[1+ (h/d)\tan i\cos\phi]^{-2}$
ranges between 0.65 and 1.7, so the azimuthal dependence should also be quite
strong.  For a dark thick disk of similar mass to the observed thin disk
and of scale height 1400 pc (such as I have hypothesized to account for
the Milky Way lensing events, Gould 1994), the optical depth would be
$8\times 10^{-7}e^{-r/d}$
along the major axis and would be a factor 30 higher along the far minor
axis and a factor 0.30 lower along the near minor axis.   Thus these
two possible disk-like structures could be easily distinguished.
As a practical matter, all three parameters would be fit
simultaneously from the optical depths as measured over the whole M31 disk.
The inner parts of the far minor axis would have to be excluded from
the fit because of contamination from the M31 bulge.

\chapter{Practical Requirements}

	The precision of the measurement described in the previous section
rests fundamentally on the possibility of
obtaining a statistically significant sample of lensing events, and hence
on observing a large number of M31 stars.  The MACHO group finds for both
the LMC bar fields and for Galactic bulge fields, that it is possible to
observe stars down to a crowding
limit of $10^6\,\rm stars\,deg^{-2}$, in $\sim 2''$ seeing.  In principle,
it is possible to reach this crowding limit in any field
by taking long enough exposures.  That is, $n$, the number of observable
stars per square degree should be related to the size of the seeing disk,
$\theta_s$, by
$$n \sim 0.3\,\theta_s^{-2} =
10^6 \biggl({2''\over \theta_s}\biggr)^2{\rm deg}^{-2}.\eqn\nexpt$$
In practice, the coefficient of $\theta_s^{-2}$ will vary from field to
field depending on the density of the partially resolved stars that are
being monitored relative
to the density of unresolved stars 1 or 2 mag below this
limit.  That is, if the crowding limit occurs at a flux level
where there are many unresolved stars within 1 mag then it will be
somewhat more difficult to follow lensing events than would be the case
if the crowding limit occurred where there are few such stars.  However, since
Tomany \& Crotts (1994) have demonstrated that there is no qualitative
barrier to finding variables in extremely crowded fields, I will assume
that equation \nexpt\ holds while recognizing that it will in general
require some adjustment.

	The total number of events observed is given by
$$N = {2\over \pi}\omega T \int d\Omega n(\Omega)\tau(\Omega),\eqn\ndef$$
where $\omega^{-1}$ is the (harmonic mean) time scale of the events,
$T$ is the length
of the observations, and $\Omega$ is the position on the sky.
Using this equation together with equation \tauexptwo,
I find a total lensing rate for the entire disk to be
$$N \sim 0.3 \omega T
\biggl({d\over \theta_s D_{\rm M31}}\biggr)^2\, {4\pi G\Sigma_0 h\over c^2}
\,\csc 2i\sim 0.5\,\omega T\biggl({ 0.\hskip-2pt ''5\over \theta_s}
\biggr)^2,\eqn\nestim$$
where $D_{\rm M31}$ is the distance to M31, and where I have ignored
a slight enhancement, $[1 - (h/d)^2\tan^2 i]^{-1/2}$, from the azimuthal
integration.  The fraction of events inside radius $r$ would be
$1 -(1 + r/d)\exp(-r/d)$, that is $\sim 70\%$ inside the two scale lengths.

	The expected time scale $\omega^{-1}$ can be estimated as follows.
Typically $\dls\sim h\sec i\sim 1500\,$pc.  For a typical disk star,
$M\sim 0.2 M_\odot$, the Einstein radius is $r_e\sim 1.5\,$A.U.  Adopting
$75\,\kms$ for a typical relative transverse speed in the inner part of
the disk, I find $\omega\sim 9{\rm yr}^{-1}$.  Thus, from equation \nestim,
observations
in $0.\hskip-2pt ''5$ seeing would yield about 5 events per year.

	The  crowding limit can be reached in the inner part of M31 with short
exposures (Tomany \& Crotts 1994).
The area within 13 kpc of the center of M31 covers only $\sim 0.8$ square
degrees (compared with 40 square degrees covered by MACHO).
Since the events are expected to last a month or more, observations
need be made only once every several days.  Hence, nearly year-round
observations are practical and do not require enormous amounts of telescope
time.  On the other hand, a substantial
fraction of the events would be in regions with significant contamination by
bulge lenses.  Thus, while a pilot ground-based program could yield some
information about the mass distribution of the M31 disk, better seeing
would be required to obtain a detailed picture.  For example, space-based
observations with $0.\hskip-2pt ''1$ seeing would yield $\sim 80$ events
over the inner two scale lengths.

\chapter{Hypothetical Observations With HST}

	As an example of a space-based program of observations, consider what
could be done with the Wide Field Camera (WFC) on the repaired {\it Hubble
Space Telescope (HST)}.  The WFC covers 4.4 square arcmin.
Therefore $\sim 2300$ square arcmin of the prime 2800 square
arcmin field inside two scale lengths
could be observed with $\sim 500$ exposures.  Although, the point
spread function of HST is $\lsim 0.\hskip-2pt ''1$, the WFC is undersampled.
I therefore adopt $\theta_s = 0.\hskip-2pt ''1$, corresponding to
$\sim 7.5\,{\rm stars}\ {\rm arcsec}^{-2}$.  I estimate the density of giant
stars with $M_I\lsim 1.2$ by assuming there are $0.01$ such stars per blue
solar luminosity and that half of these stars lie behind the plane and are
blocked by dust.
I then find $\sim 0.3\,{\rm pc}^{-2}\,\exp(-r/d)$
or $\sim 15\,{\rm arcsec}^{-2}\,\exp(-r/d)$.  Thus, to achieve the crowding
limit at 1 scale length, the observations must be sensitive to $M_I\sim 1.2$
or $I\sim 25.8$.  While the new WFC camera has not yet been calibrated, I
extrapolate from experience with the old camera to estimate that 30 minute
exposures would yield 7\% relative photometric errors.  Hence observations
 would require $\sim 250$ hours of telescope time, implying that the
observations could not be carried out once per week even if the telescope
were devoted to this project full time.

	This calculation shows that {\it HST} is not suitable for this
project.  A specially designed telescope with a much larger field of
view would be required.

\chapter{Lensing By a Spheroid}

	As discussed by Giudice et al.\ (1994) it is possible that the
MACHO and EROS events toward the LMC are generated by low-mass stars
in the Galactic
spheroid.  If so, it would also be plausible to expect significant lensing
by the M31 spheroid.  Here, I present analytic results for lensing of
sources in the M31 disk by a spherical spheroid having core radius
$a$ and  asymptotic fall-off as $r^{-n}$, $n>2$.  I then
discuss the implications for lensing experiments toward M31.

\section{Analytic Formulae}

	Suppose that lenses are distributed in the M31 spheroid with
a density distribution
$$\rho(r) = {\rho_0\over [1 + (r/a)]^{n/2}},\eqn\rhospher$$
with $n>2$.  The Einstein ring radius for sources in the disk
is again given by equation \redef.  As in the case of lensing by disk
stars, $\dol\sim \dos$, so that the Einstein ring reduces to equation
\redeftw.  Hence the optical depth is given by
$$\tau = {4\pi G\rho_0\over c^2}\int_{-q}^\infty d\ell\,{a^n(\ell + q)\over
(a^2+b^2 + \ell^2)^{n/2}},\eqn\tauone$$
where $b$ is the impact parameter, $q$ is the distance from the disk star
to the point of nearest approach of the line of sight
to the center of M31, and $\ell$
parameterizes the distance along the line of sight.  This becomes
$$\tau = {4\pi G\rho_0 a^2\over c^2}\,\biggl(1 + {b^2\over a^2}
\biggr)^{1-{n\over 2}}\biggl(
{\cos^{n-2}\theta_0\over n-2} +
\tan\theta_0\int_{-\theta_0}^{\pi/2}d\theta\,\cos^{n-2}\theta
\biggr), \eqn\tautwo$$
where
$$\tan \theta_0\equiv {q\over \sqrt{a^2+ b^2}}.\eqn\thetazero$$
Thus, along the major axis
$$\tau = {4\pi G\rho_0 a^2\over (n-2)c^2}\,
\biggl(1+ {b^2\over a^2}\biggr)^{1-{n\over 2}}\qquad
{\rm (major}\ {\rm axis)} \eqn\taumajor$$
For two special cases, the results are
$$\tau = {4\pi G\rho_0 a^3\over c^2(a^2+b^2)}\,
(q + \sqrt{a^2+b^2+q^2}), \qquad (n=3),\eqn\taunthree$$
and
$$\tau = {2\pi G\rho_0 a^4\over c^2(a^2+b^2)}\,
\biggl[1 + {q\over\sqrt{a^2+b^2}}
\biggl({\pi\over 2} + \tan^{-1}{q\over\sqrt{a^2+b^2}}\biggr)\biggr],
\qquad (n=4).\eqn\taunfour$$

\section{Implications For Observations of M31}

	The pattern of the optical depth for a power-law
spheroid differs substantially from that of an exponential disk.  The
optical depth of a disk is exponential in radius, while the optical
depth of a spheroid is a power law with an index two lower than the
spheroid itself.  This is immediately apparent for the major axis
from equation \taumajor, but in fact is true at any fixed azimuthal angle.
The azimuthal dependence of the optical depth is
more pronounced for a spheroid than for a disk.  For $a\ll b$, the ratio on
the (far minor axis):(major axis):(near minor axis) at fixed radius is
0.51:1:29 for an $n=3$ spheroid and 0.34:1:175 for $n=4$.  For a spheroid
with a steep index, it is possible that at a given radius,
the near minor axis will be dominated by the spheroid, while the major
axes are dominated by the disk.

{\bf Acknowledgements:  } I would like to thank Chris Flynn for making
several helpful suggestions.
\endpage

\Ref\Alcock{Alcock, C., et al.\ 1993, Nature, 365, 621}
\Ref\Aubourg{Aubourg, E., et al.\ 1993, Nature, 365, 623}
\Ref\BT{Binney, J.\ \& Tremaine, S.\ 1987, Galactic Dynamics, (Princeton:
Princeton Univ.\ Press)}
\Ref\apsc{Crotts, A.\ P.\ S.\ 1992, ApJ, 399, L43}
\Ref\gmr{Giudice, G.\ F., Mollerach, S., \& Roulet, E.\ 1994, Phys Rev D,
submitted}
\Ref\gtwo{Gould, A.\ 1994, ApJ, 421, L71}
\Ref\gmb{Gould, A., Miralda-Escud\'e, \& Bahcall, J.\ N.\ 1994,
ApJ, 423, L105}
\Ref\grie{Griest, K.\ et al.\ 1991, ApJ, 372, L79}
\Ref\Pac{Paczy\'nski, B.\ 1986, ApJ, 304, 1}
\Ref\Pac{Paczy\'nski, B.\ 1991, ApJ, 371, L63}
\Ref\tc{Tomany, A.\ B.\ \& Crotts, A.\ P.\ S.\ 1994, ApJ Letters, submitted}
\Ref\ogle{Udalski, A., Szyma\'nski, J.,  Kaluzny, J., Kubiak, M.,
Krzemi\'nski, W., Mateo, M., Preston, G.\ W., \&
Paczy\'nski, B. 1993, Acta Astronomica, 43, 289}
\Ref\vdK{van der Kruit, P.\ C. 1989, The Milky Way As A Galaxy,
R.\ Buser and I.\ R.\ King, eds., (Mill Valley: University Science Books)}
\refout
\endpage
\endpage
\bye